\documentclass[11pt]{article}
\usepackage{moriond,epsfig}

\def\NIMA{{\em Nucl. Instrum. Methods} A}

\def\PLB{{\em Phys. Lett.}  B}
\def\PRL{{\em Phys. Rev. Lett.}}
\def\PRD{{\em Phys. Rev.} D}

\def\be{\begin{equation}}
\def\ee{\end{equation}}
\def\bea{\begin{eqnarray}}
\def\eea{\end{eqnarray}}

\begin{document}
\vspace*{3.5cm}
\title{PHENIX SPIN PROGRAM, RECENT RESULTS}

\author{ A.~BAZILEVSKY, for the PHENIX Collaboration }

\address{Brookhaven National Laboratory, Upton, NY, USA}

\maketitle\abstracts{
Acceleration of polarized protons in the Relativistic Heavy Ion Collider 
(RHIC) provides a unique tool to study the spin structure of the nucleon. 
We give a brief overview of the PHENIX program to investigate the unknown 
gluon and flavor decomposed sea quark polarization in the proton,
utilizing polarized proton collisions at RHIC.
We report first results from the PHENIX experiment on transverse single-spin 
asymmetry in $\pi^0$ and charged hadron production and longitudinal 
double-spin asymmetry in $\pi^0$ production, at mid-rapidity. 
}

\section{Introduction}

Spin is one of the most fundamental properties of the elementary particles. 
The spin of the proton had been believed to be carried by the valence quarks 
for long years. From polarized deep inelastic (DIS) lepton-nucleon scattering 
experiments over the past 20 years it has been found that on average little 
of the proton spin is carried by quarks and antiquarks~\cite{DISsigma}.  
Therefore, most of the proton spin must be carried by the gluons and orbital 
angular momentum. It may also indicate that the sea quark polarization 
is large and anti-parallel to the proton spin. 

DIS experiments have constrained the possible
gluon polarization in the proton through the measurement of scaling violation
in inclusive polarized scattering~\cite{DISg1}, and through semi-inclusive 
measurements of two hadrons to isolate the photon-gluon fusion 
diagram~\cite{DIShad}.  However, the reach of these
measurements has been limited, due to the low energy available for fixed target
experiments, and due to luminosity limitations.  
A fixed target experiment at Fermilab first presented a measurement with 
strongly interacting probes~\cite{E704all}, again with limited sensitivity 
due to the fixed target energy and available luminosity.  
At this time, the gluon contribution to the proton spin is largely unknown.

The colliding polarized proton beams at the Relativistic Heavy Ion Collider 
provide a new laboratory to probe the proton spin structure with
strongly interacting probes. There are several processes where gluons 
participate directly, such as prompt photon and heavy quark production. 
Flavor decomposition of quark and antiquark polarization can be done 
using $W$ production. These measurements use longitudinal beam polarization. 

Surprisingly large transverse single-spin asymmetries ($A_N$) 
have been observed in a number of experiments \cite{E704,hermes,star}, 
ranging in energy $\sqrt{s}$=20--200 GeV. 
PHENIX will further investigate the origin of such asymmetries, 
colliding transversely polarized protons.

\section{PHENIX spin program, recent results}

\subsection{Gluon polarization}

Measurement of the gluon polarization ($\Delta g$) in a polarized proton is 
a major goal of the PHENIX spin program. The main probes are high $p_T$ 
prompt photon production, jet (or leading hadron) production and heavy-flavor 
production. 

The PHENIX experiment  \cite{phenix_nim} has reported the unpolarized cross 
section for $\pi^0$ and prompt photon production in $pp$ collisions at 
midrapidity, which is described well by next-to-leading order perturbative 
QCD (NLO pQCD) calculations \cite{pi0_prl,gamma_okada}. In pQCD the 
longitudinal double spin asymmetry $A_{LL}$ for those processes is directly 
sensitive to the polarized gluon distribution function in the proton 
\cite{theory,theory_jet,frixione}. $A_{LL}$ is defined:
\begin{equation}
A_{LL}=\frac{\sigma_{++}-\sigma_{+-}}{\sigma_{++}+\sigma_{+-}}
\label{eq:a_ll}
\end{equation}
where $\sigma_{++}$ ($\sigma_{+-}$) is the cross section of the reaction
when two colliding particles have the same (opposite) helicity. 
From Eq.~\ref{eq:a_ll} and equating the cross section to the ratio 
of experimental yield ($N$) and the integrated luminosity ($L$), 
$A_{LL}$ is expressed as 
\begin{equation}
A_{LL} = \frac{1}{|P_{B}P_{Y}|} \cdot \frac{N_{++}-R_{LL}\cdot
N_{+-}}{N_{++} + R_{LL} \cdot N_{+-}};~~~R_{LL}=\frac{L_{++}}{L_{+-}},
\label{eq:A_LL}
\end{equation}
where $P_{Y(B)}$ are the polarizations of the RHIC ``yellow'' (``blue'') 
beams. 

\begin{figure}[]
\vspace{-0.5cm}
\begin{minipage}[t]{90mm}
\psfig{figure=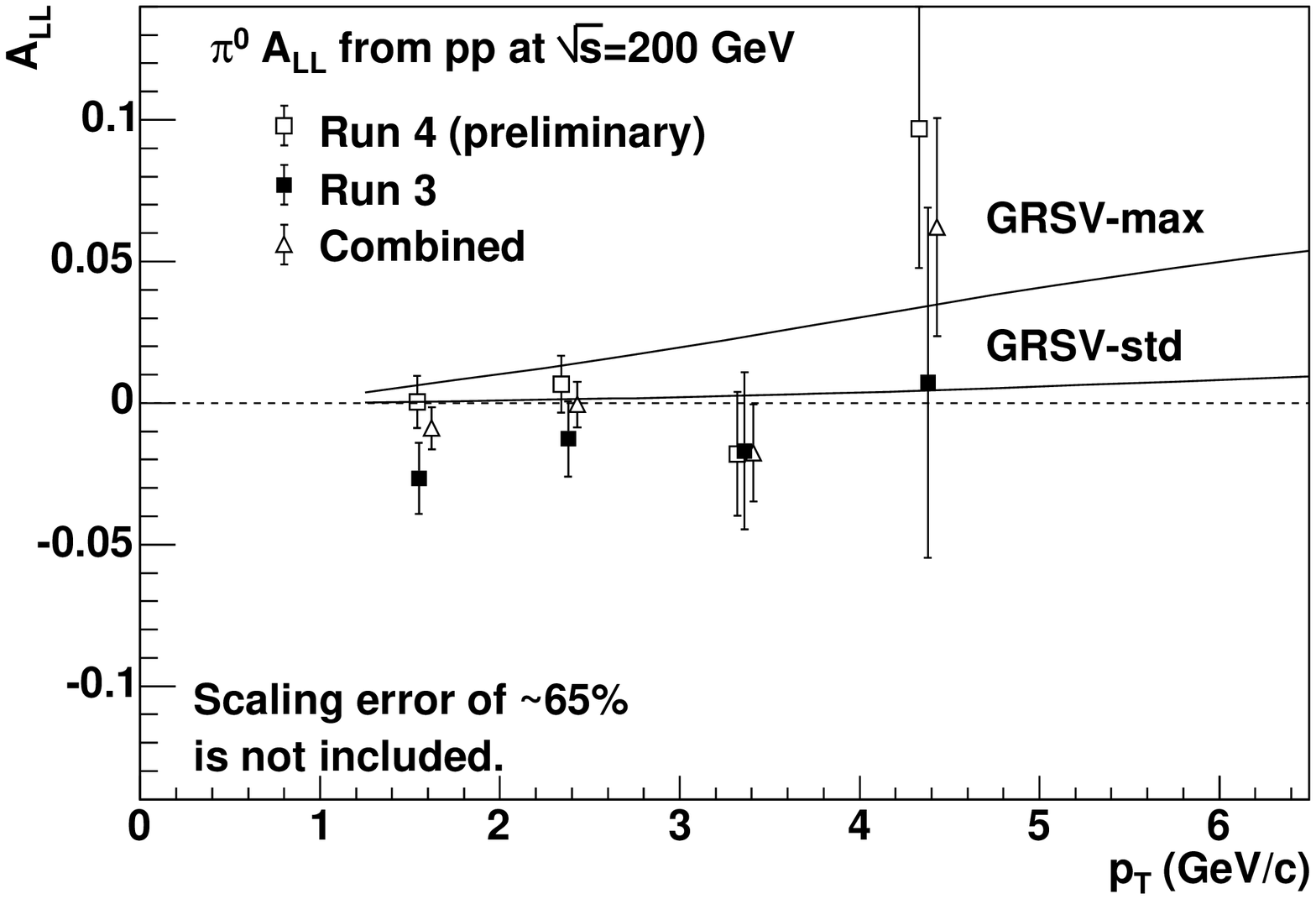,height=55mm}
\end{minipage}
\begin{minipage}[t]{90mm}
\psfig{figure=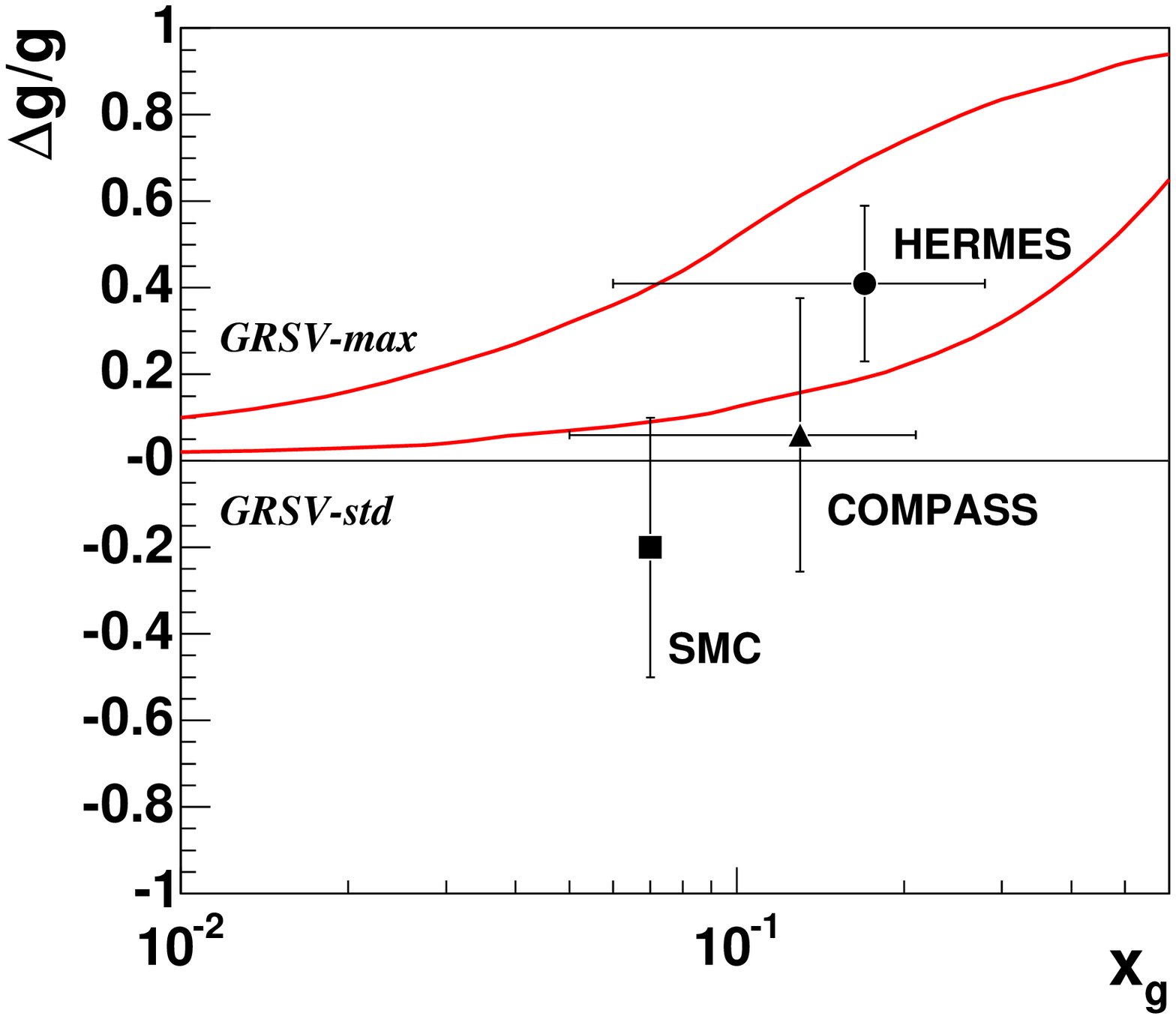,height=60mm}
\end{minipage}
\caption{\label{fig:all} Left: $A_{LL}^{\pi^{0}}$ 
versus mean $p_{T}$ of $\pi^0$'s in each bin; 
a scale uncertainty of $\pm65\%$ is not included.
Two theoretical calculations based on
NLO pQCD are also shown for comparison with the data. 
Right: HERMES, SMC and COMPASS results compared  
to the same theoretical calculations.}
\end{figure}

PHENIX has already presented and published first $A_{LL}$ data for 
inclusive $\pi^0$ production from the 2003/2004 RHIC runs 
(Run3/Run4) \cite{all_prl}. The results are shown on Fig.~\ref{fig:all}(left). 
Two theoretical curves based on NLO pQCD represent different assumptions 
for the gluon polarization~\cite{theory,GRSV}. 
The gluon polarization
contributes to $A_{LL}$ through gluon-gluon and gluon-quark subprocesses,
with the gluon-gluon contribution significantly 
larger at mid-rapidity and for the estimated gluon momentum fraction 
for these results $x \approx 0.03-0.1$~\cite{theory_jet}.
The results are consistent with zero or small gluon
polarization, with a confidence level (CL) of 21--24\% for GRSV-std, for 
the range in polarization scale uncertainty of the measurement. 
The results are less consistent with a large gluon polarization, 
with CL=0--6\% for GRSV-max. 

Fig.~\ref{fig:all}(right) compares results obtained by HERMES, SMC and 
COMPASS collaborations to the same theoretical models. They constrain the 
possible gluon polarization in the proton through semi-inclusive
measurements of two hadrons to utilize the photon-gluon fusion
process~\cite{DIShad}. HERMES result sensitivity in its kinematic 
range is not yet enough to distinguish between GRSV-std and GRSV-max models, 
while SMC (probing the same $x_g$ as PHENIX) and COMPASS points, 
similar to the PHENIX result, favor more the GRSV-std calculation. 

Even with a small integrated luminosity (220 nb$^{-1}$ in Run3 and 
75 nb$^{-1}$ in Run4) and moderate beam polarization ($\sim$27\% in Run3 
and $\sim$40\% in Run4) PHENIX data have already started constraining 
$\Delta g$ at a level comparable to polarized-DIS data. The increased 
integrated luminosity and improved polarization expected for the 
on-going Run5 should provide an order of magnitude decrease in statistical 
uncertainties. 

Prompt photons ($\gamma^{prompt}$) are considered as the 
``golden'' probe for $\Delta g$ due to the dominance of Gluon Compton 
graph in $\gamma^{prompt}$ production \cite{frixione}. Measurement of 
$A_{LL}(\gamma^{prompt})$ requires considerable integrated luminosity. 
The results from 70\% polarized proton beam collisions corresponding to 
$\sim 200~pb^{-1}$ are expected in 2007-2008. 

\subsection{Antiquark polarization}

Detailed flavor analysis is possible with $W^{\pm}$ production measurements, 
since the participating quark and antiquark helicities are fixed and 
the dominant contributors are $u$, $d$, $\bar{u}$ and $\bar{d}$: 
$u\bar{d} \rightarrow W^{+}$, $d\bar{u} \rightarrow W^{-}$. 
The parity violating longitudinal single-spin asymmetry is 
\begin{equation}
A_{L} = -\frac{\sigma_{+}-\sigma_{-}}{\sigma_{+}+\sigma_{-}}, 
%\label{eq:a_ll}
\end{equation}
where $\sigma_{+}$ ($\sigma_{-}$) represents a cross section of the reaction 
with the initial proton helicity ``+'' (``$-$''), protons from the other 
beam being unpolarized (or helicity states averaged). 
For $W^{-}$ production 
\begin{equation}
A^{W^{-}}_{L} = \frac{\Delta d(x_1) \bar{u}(x_2) - \Delta \bar{u}(x_1) d(x_2)}
{d(x_1) \bar{u}(x_2) + \bar{u}(x_1) d(x_2)}.
\label{eq:al_w}
\end{equation}
To obtain the asymmetry for $W^{+}$, one has to interchange $u$ and $d$ 
in Eq.~\ref{eq:al_w}.
At large $x_2$, where $\bar{u}(x_2)$ is small, $A^{W^-}_{L}$ becomes 
$-\Delta\bar{u}/\bar{u}$, and at large $x_1$, $A^{W^-}_{L}$ is 
$\Delta d/d$. Likewise from $W^+$, $\Delta\bar{d}/\bar{d}$ and 
$\Delta u/u$ can be extracted.

In the central arms $W^{\pm}$ are identified as a Jacobian peak in the $p_T$ 
spectrum of electrons and positrons. In the muon arms the $p_T$ spectrum of 
muons is predominantly from $W^{\pm}$ decays for the $p_T$ region 
above 20 GeV/c. First results on $A_L^{W^{\pm}}$ from longitudinally 
polarized $pp$ collisions at $\sqrt{s}$=500 GeV are expected in 2009-2010. 

\subsection{Transverse spin}

Exciting physics prospects also arise for transverse polarization of the 
RHIC proton beams. One is the possibility of a first measurement of the 
quark transversity densities $\delta q$. Difference between 
$\Delta q$ and $\delta q$ provide a measure of the relativistic nature 
of quarks inside the nucleon. 

Studies of transverse single-spin asymmetries ($A_N$) is a further interesting 
application. Similarly to $A_{LL}$ (Eq.~\ref{eq:A_LL}), $A_N$ is 
experimentally defined: 
\begin{equation}
A_{N} = \frac{1}{|P|} \cdot \frac{1}{\langle |cos\phi| \rangle} \cdot 
\frac{N_{\uparrow}-R_{N} \cdot N_{\downarrow}}
{N_{\uparrow}+R_{N} \cdot N_{\downarrow}};~~~R_{N}=\frac{L_{\uparrow}}{L_{\downarrow}},
\label{eq:A_N}
\end{equation}
where we compare particle production on the left side of upward and downward 
polarized beam; 
$\langle |cos\phi| \rangle$ is the average $|cos\phi|$ of the detected 
particles, with $\phi$ the azimuthal angle.

Over the years, a number of models based on pQCD have 
been developed to explain the asymmetries observed at lower energies. 
Among them are the Sivers effect \cite{sivers}, transversity and the Collins 
effect \cite{collins}, and various models which attribute the observed 
asymmetries to higher twist contributions (e.g. \cite{htwist}). 

\begin{figure}
\begin{center}
\vspace{-0.5cm}
\includegraphics[height=13pc]{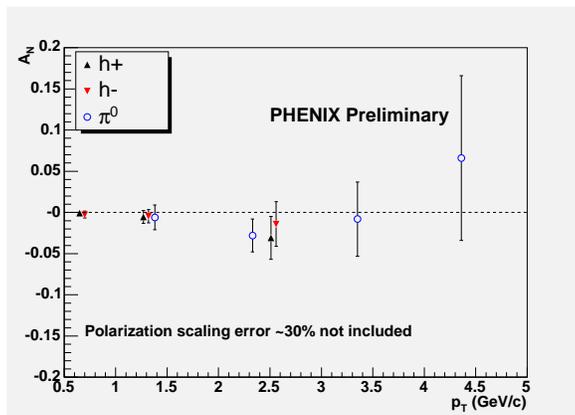}
\caption{\label{fig:an} Transverse single spin asymmetry for inclusive 
charged hadrons and neutral pions; a scale uncertainty of $\pm30\%$ is 
not included. 
}
\end{center}
\end{figure}

Fig.~\ref{fig:an} shows the $A_{N}$ in $\pi^0$ and charged hadron 
production vs $p_T$. The transverse single-spin asymmetries are consistent 
with zero over the measured transverse momentum range.  
A small or zero asymmetry in this kinematic region follows the trend 
of previous results, which indicate a decreasing asymmetry at decreasing 
$x_{F}$~\cite{E704Central,star}.

\section*{Acknowledgments}
We thank the staff of the Collider-Accelerator Department, Magnet
Division, and Physics Department at BNL and the RHIC polarimetry group for
their vital contributions.  We thank W.~Vogelsang for informative
discussions.  We acknowledge support from the Department of Energy and NSF
(U.S.A.), MEXT and JSPS (Japan), CNPq and FAPESP (Brazil), NSFC (China),
IN2P3/CNRS, CEA, and ARMINES (France), BMBF, DAAD, and AvH (Germany), OTKA
(Hungary), DAE and DST (India), ISF (Israel), KRF and CHEP (Korea), RAS,
RMAE, and RMS (Russia), VR and KAW (Sweden), U.S. CRDF for the FSU,
US-Hungarian NSF-OTKA-MTA, and US-Israel BSF.

\end{document}